\documentclass{midl} % Include author names
\jmlrproceedings{MIDL}{Medical Imaging with Deep Learning}
\jmlrpages{}
\jmlryear{2022}

\jmlrworkshop{Short Paper -- MIDL 2022 submission}
\jmlrvolume{-- Under Review}
\editors{Under Review for MIDL 2022}

\title[Dual Branch Prior-SegNet]{Dual Branch Prior-SegNet: CNN for Interventional CBCT using Planning Scan and Auxiliary Segmentation Loss}

\midlauthor{\Name{Philipp Ernst\nametag{$^{1,2}$}}\Email{philipp.ernst@ovgu.de}\AND
\Name{Suhita Ghosh\nametag{$^{1}$}}\Email{suhita.ghosh@ovgu.de}\AND
\Name{Georg Rose\nametag{$^{2,3}$}}\Email{georg.rose@ovgu.de}\AND
\Name{Andreas N\"urnberger\nametag{$^{1}$}}\Email{andreas.nuernberger@ovgu.de}\\
\addr $^{1}$ Faculty of Computer Science, University of Magdeburg, Germany\\
\addr $^{2}$ Research Campus \textit{STIMULATE}, University of Magdeburg, Germany\\
\addr $^{3}$ Institute for Medical Engineering, University of Magdeburg, Germany
}

\begin{document}

\maketitle

\begin{abstract}
This paper proposes an extension to the Dual Branch Prior-Net for sparse view interventional CBCT reconstruction incorporating a high quality planning scan. An additional head learns to segment interventional instruments and thus guides the reconstruction task. The prior scans are misaligned by up to \textpm 5deg in-plane during training. Experiments show that the proposed model, Dual Branch Prior-SegNet, significantly outperforms any other evaluated model by \textgreater2.8dB PSNR. It also stays robust wrt. rotations of up to \textpm5.5deg.
\end{abstract}


\section{Introduction}
Reduction of X-ray exposure during CT-guided medical interventions is essential for surgeons and patients not to develop harmful diseases, but results in lower quality reconstructions compared to using full dose. \citet{ghosh2022} presented a CNN incorporating a high quality planning scan for the reconstruction of sparse view interventional CBCT. We propose to add a segmentation head parallel to the reconstruction head, segmenting interventional instruments to guide the reconstruction task. The main contributions are: (i) evaluating the performance with the segmentation head and (ii) determining the limits of prior scan misalignment for in-plane rotations.

\section{Methods}
% \subsection{Dual Branch Prior-SegNet}
The network is a multi-scale dual branch CNN extracting features from both a sparse view interventional CBCT scan and a high quality planning scan separately. These are combined via skip connections in the decoding path. After the last upsampling, we add another convolutional block with sigmoid activation parallel to the reconstruction block of the original network to segment interventional instruments. The hypothesis is that giving the network the additional task of segmentation increases the quality of the reconstruction since it is forced to learn what causes the most prominent streaking artifacts.
% \subsection{Loss Function}
The loss function is a combination of MSE reconstruction and Dice segmentation loss:
$L(p, g) = L_{MSE}(p_r, g_r) + \lambda\cdot L_{Dice}(p_s, g_s)$
for the prediction $p$ and the ground truth $g$, $r$ and $s$ denoting reconstruction and segmentation. $\lambda$ was set to 1e-3 empirically.
% \subsection{Data Set and Preprocessing}
Training data was simulated by combining LungCT-Diagnosis~\cite{lungct-diagnosis} volumes and in-house needle scans. Segmentations were created by thresholding the needle scans at 900HU. The data was normalized by dividing the attenuation coefficients by the 99th percentile. 13 equiangular projections on a circular trajectory were simulated.
% \subsection{Training Details}
The networks were trained for 150 epochs using Adam (lr=1e-3) with a batch size of 32 and mixed precision. Online augmentations were performed including random rotations, scalings and flips and up to \textpm 5deg in-plane misalignment of the prior planning scan.

\section{Results}
\begin{table}
 % The first argument is the label.
 % The caption goes in the second argument, and the table contents
 % go in the third argument.
\floatconts
  {tab:metrics}%
  {\caption{Mean and standard deviation over all axial slices of the test set without misalignment of the planning scan.
%   \textit{(no mis)} indicates models trained without misalignment of the prior.
  }}%
  {\begin{tabular}{lccc}
  \bfseries Model/Method & \bfseries SSIM [\%] & \bfseries PSNR [dB] & \bfseries RMSE [HU]\\
  FDK & 17.47\textpm 7.35 & 10.33\textpm 1.48 & 1473.60\textpm 349.98\\
  FDKConvNet & 64.16\textpm 11.18 & 21.78\textpm 2.38 & 234.40\textpm 67.38\\
  Dual Branch Prior-Net & 96.71\textpm 3.31 & 41.09\textpm 3.50 & 28.70\textpm 15.09\\
  Dual Branch Prior-SegNet & 97.15\textpm 3.47 & 43.97\textpm 4.95 & 23.21\textpm 18.11\\
  \end{tabular}}
\end{table}
\tableref{tab:metrics} shows the results of the different models evaluated on the test set (without misalignment of the prior scan). All models outperform the direct sparse view FDK reconstruction by a large margin, while the Dual Branch models further increase the quality noticeably compared to FDKConvNet~\cite{jin2017}. 
% The models trained with augmentations and misalignments do not clearly increase the SSIM but have a positive effect on PSNR and RMSE values. 
Prior-SegNet results in the lowest errors. Wilcoxon signed-rank tests reveal that the proposed model significantly outperforms any other model pair-wise (p-value $<0.5\%$).
\begin{figure}[htbp]
 % Caption and label go in the first argument and the figure contents
 % go in the second argument
\floatconts
  {fig:example}
  {\caption{a) Exemplary ROI around needle of different models/methods. b) Reconstruction errors using misaligned (rotated) prior scan (median and interquartile range).}}
  {\includegraphics[width=\linewidth]{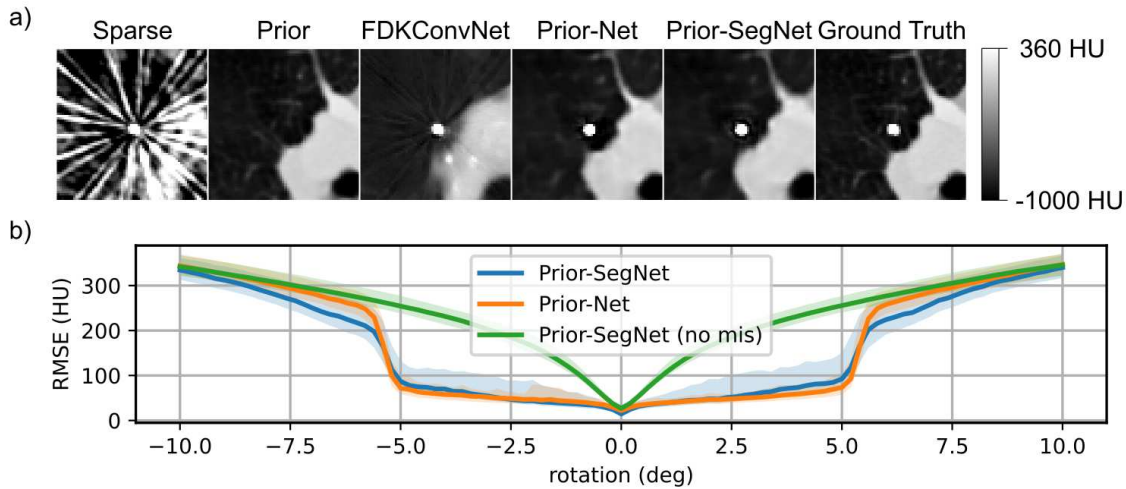}}
\end{figure}
\figureref{fig:example} a) shows a region of interest centered around the needle in the axial slice with the highest errors of the first test subject with a misalignment of the prior scan by 2deg. FDKConvNet cannot recover fine structures and inserts ghosting artifacts of the needle. Prior-Net slightly blurs a small region around the needle whereas Prior-SegNet inserts a slight halo. Both compensate for the misalignment of the prior.
\figureref{fig:example} b) shows the errors wrt. rotated prior scans. Prior-SegNet (no mis) was trained without misalignment of the prior scan and performs worst. Prior-Net performs best for $2.5\text{deg}\leq|\alpha|\leq5.5\text{deg}$ and Prior-SegNet for all other angles.

\section{Discussion and Conclusion}
Incorporating a prior planning scan for sparse view interventional CBCT is a simple way to increase the quality of the reconstructions. FDKConvNet has no further prior information and barely reconstructs tissues that are not significantly affected by streaking artifacts. Training with misaligned priors is essential to keep the quality at a high level. Though not trained with rotations $|\alpha|>5\text{deg}$, Prior-Net and Prior-SegNet compensate for up to \textpm5.5deg. Confirming the initial hypothesis, the segmentation head facilitates the reconstruction task for small angles and seems to generalize better for high angles. Future work will focus on different types of misalignment, e.g. translation and elastic distortions, and evaluating their limits systematically. Moreover, this preliminary work is based on simulations and has to be evaluated for real CBCT data.
The code is available on Github\footnote{\url{https://github.com/phernst/prior-segnet}}.

% Acknowledgments---Will not appear in anonymized version
\midlacknowledgments{This work was supported by the ESF (project no. ZS/2016/08/80646).}

\bibliography{midl-shortpaper}

\end{document}